\documentclass[12pt]{article}
\usepackage{graphicx}\usepackage{qcmc06}
\usepackage{theorem,amsmath,amssymb,mathrsfs}
\def\<{\langle}\def\>{\rangle}
\def\:{\hbox{\bf :}}
\renewcommand{\leq}{\leqslant}
\def\Reals{\mathbb R}\def\Cmplx{\mathbb C}
\def\map#1{{\mathscr{#1}}}

\def\set#1{{\sf #1}}\def\alg#1{{\mathcal #1}}\def\aA{\alg{A}}
\def\aI{\alg{I}}

\def\sH{\set{H}}

\def\ie{i. e. }
\def\n#1{|\!|#1|\!|}
\def\dag{\dagger}\def\eff{{\rm eff}}
 
\newtheorem{postulate}{Postulate}

\def\trnsfrm#1{\mathscr #1}
\def\tA{\trnsfrm A}\def\tB{\trnsfrm B}\def\tC{\trnsfrm C}
\def\tS{\trnsfrm S}
\def\tI{\trnsfrm I}\def\tT{\trnsfrm T}\def\tU{\trnsfrm U}\def\tX{\trnsfrm X}

\def\AA{\mathbb A}\def\AB{\mathbb B}\def\AL{\mathbb L}
\def\cA{{\underline{\tA}}}\def\cB{\underline{\tB}}\def\cC{\underline{\tC}}\def\cT{\underline{\tT}}

\def\Stset{{\mathfrak S}}
\def\Trnset{{\mathfrak T}}
\def\Cntset{{\mathfrak P}}
\begin{document}
\pagestyle{empty}
\title{Operational axioms for a C${}^*$-algebraic formulation of Quantum Mechanics}
\author{Giacomo Mauro D'Ariano\refnote{1}}
\affiliation{\affnote{1}Dipartimento di Fisica ``A. Volta'', via Bassi 6, 27100 Pavia, Italy}
\begin{abstract}
  A C${}^*$-algebra formulation of Quantum Mechanics is derived from purely operational axioms in
  which the primary role is played by the {\em transformations} that the system undergoes in the
  course of an {\em experiment}. The notion of the {\em adjoint} of a transformation is based on the
  postulated existence of {\em faithful states} that allows one to calibrate the experimental
  apparatus.  
\end{abstract}
\section{Introduction}
In a set of recent papers \cite{darianoVax2006,dariano-beyond} I recently showed how it is possible
to derive the mathematical formulation of Quantum Mechanics in terms of complex Hilbert spaces or in
terms of C${}^*$-algebras, starting from five purely operational Postulates concerning {\em
  experimental accessibility and simplicity}. The starting point for the axiomatization is a seminal
definition of {\em physical experiment} which entails the thorough series of notions that are at the
basis of the axiomatization. In the present short account I will briefly review the derivation of a
C${}^*$-algebra formulation from only two Postulates on the physical experiment, based on a
operational notion of the {\em adjoint} of a transformation, which follows from the postulated
existence of {\em faithful states}. Such states are crucial for calibrating the experimental
apparatus, and their basic idea comes from modern Quantum Tomography \cite{tomo_lecture,calib}.  The
quantum C${}^*$-algebra representation of the transformations (for generally infinite dimensions) is
then derived from the Postulates via a Gelfand-Naimark-Segal (GNS) construction\cite{GNS}.
\section{The postulates} 
The general premise of the present axiomatization is the fact that one performs experiments to get
information on the {\em state} of an {\em object} physical system, and the knowledge of such a state
will then enable to predict the results of forthcoming experiments. Moreover, since we necessarily
work with only partial {\em a priori} knowledge of both system and experimental apparatus, the rules
for the experiment must be given in a probabilistic setting. {\em What we mean by experiment?}  An
experiment on an object system consists in making it interact with an apparatus: the interaction
between object and apparatus produces one of a set of possible transformations of the object, each
one occurring with some probability. Information on the state of the object at the beginning of the
experiment is gained from the knowledge of which transformation occurred, which is the "outcome" of
the experiment signaled by the apparatus.
\par We can now introduce the two postulates.
\begin{postulate}[Independent systems]\label{p:independent} There exist independent physical systems.
\end{postulate}
\begin{postulate}[Symmetric faithful state]\label{p:faith} For every composite system made of two identical
  physical systems there exist a symmetric joint state that is both dynamically and preparationally
  faithful.
\end{postulate}
\medskip

\section{The statistical and dynamical structure}
The starting point of the axiomatization is the identification {\bf experiment }$\equiv${\em set of
  transformations} $\AA\equiv\{\tA_j\}$ that can occur on the object. The apparatus signals which
transformation actually occurs. Now, since the knowledge of the state of a physical system allows us
to predict the results of forthcoming experiments on the object, then it would allow us to evaluate
the probability of any possible transformation in any conceivable experiment. Therefore, by
definition, a {\bf state} $\omega$ of a system is a rule providing probabilities of transformation,
and $\omega(\tA)$ is the probability that the transformation $\tA$ occurs.  We clearly have the
completeness $\sum_{\tA_j\in\AA}\omega(\tA_j)=1$, and assume $\omega(\tI)=1$ for the identical
transformation $\tI$, corresponding to adopting $\tI$ as the free evolution (this is the {\em Dirac
  picture}, \ie a suitable choice of the lab reference frame). In the following for a given physical
system we will denote by $\Stset$ the set of all possible states and by $\Trnset$ the set of all
possible transformations. 
\par When composing two transformations $\tA$ and $\tB$, the probability $p(\tB|\tA)$ that $\tB$
occurs conditional on the previous occurrence of $\tA$ is given by the Bayes rule for conditional
probabilities $p(\tB|\tA)=\omega(\tB\circ\tA)/\omega(\tA)$.  This sets a new probability rule
corresponding to the notion of {\bf conditional state} $\omega_\tA$ which gives the probability that
a transformation $\tB$ occurs knowing that the transformation $\tA$ has occurred on the object in
the state $\omega$, namely $\omega_\tA\doteq\omega(\cdot\circ\tA)/\omega(\tA)$\footnote{Throughout
  the paper we will make extensive use of the functional notation with the central dot corresponding
  to a variable transformation}. We can see that the notion of ``state'' itself logically implies
the identification {\em evolution}$\equiv${\em state-conditioning}, entailing a {\em linear
  action of transformations on states} (apart from normalization)
$\tA\omega:=\omega(\cdot\circ\tA)$: this is the same concept of {\bf operation} that we have in
Quantum Mechanics, which gives the conditioning $\omega_\tA=\tA\omega/\tA\omega(\tI)$.  In other
words, this is the analogous of the Schr\"{o}dinger picture evolution of states in Quantum Mechanics
(clearly such identification of evolution as state-conditioning also includes the deterministic case
$\tU\omega=\omega(\cdot\circ\tU)$ of transformations $\tU$ with
$\omega(\tU)=1\,\forall\omega\in\Stset$---the analogous of quantum unitary evolutions and channels. 

From the Bayes conditioning it follows that we can define two complementary types of equivalences
for transformations: {\em dynamical} and {\em informational}. The transformations $\tA_1$ and
$\tA_2$ are {\bf dynamically equivalent} when $\omega_{\tA_1}=\omega_{\tA_2}$
$\forall\omega\in\Stset$, whereas they are {\bf informationally equivalent} when
$\omega(\tA_1)=\omega(\tA_2)$ $\forall\omega\in\Stset$. The two transformations are then completely
equivalent (write $\tA_1=\tA_2$) when they are both dynamically and informationally equivalent,
corresponding to the identity $\omega(\tB\circ\tA_1)=\omega(\tB\circ\tA_2)$,
$\forall\omega\in\Stset,\;\forall\tB\in\Trnset$. We call {\bf effect} the informational equivalence
class of transformations (this is the same notion introduced by Ludwig\cite{Ludwig-axI}).  In the
following we will denote effects with the underlined symbols $\cA$, $\cB$, etc., or as $[\tA]_\eff$,
and we will write $\tA_0\in\cA$ meaning that "the transformation $\tA$ belongs to the equivalence
class $\cA$", or "$\tA_0$ has effect $\cA$'', or "$\tA_0$ is informationally equivalent to $\tA$".
Since, by definition one has $\omega(\tA)\equiv\omega(\cA)$, we will legitimately write
$\omega(\cA)$ instead of $\omega(\tA)$. Similarly, one has $\omega_\tA(\tB)\equiv \omega_\tA(\cB)$,
which implies that $\omega(\tB\circ\tA)=\omega(\cB\circ\tA)$, leading to the chaining rule
$\cB\circ\tA\in\underline{\tB\circ\tA}$ corresponding to the "Heisenberg picture" evolution of
transformations acting on effects (notice how transformations act on effects from the right). Now,
by definitions effects are linear functionals over states with range $[0,1]$, and, by duality, we
have a convex structure over effects, and we will denote their convex set as $\Cntset$.  An {\bf
  observable} is just a complete set of effects $\AL=\{l_i\}$ of an experiment $\AA=\{\tA_j\}$,
namely one has $l_i=\underline{\tA_j}$ $\forall j$ (clearly, one has the completeness relation
$\sum_il_i=1$). We will call the observable $\AL=\{l_i\}$ is {\bf informationally complete} when
each effect $l$ can be written as a linear combination $l=\sum_ic_i(l)l_i.$ of elements of $\AL$,
and when these are linearly independent we will call the informationally complete observable {\em
  minimal}.

\par The fact that we necessarily work in the presence of partial knowledge about both object and
apparatus corresponds to the possibility of incomplete specification of both states and
transformations, entailing the convex structure on states and the addition rule for {\em coexistent
  transformations}, namely for transformations $\tA_1$ and $\tA_2$ for which
$\omega(\tA_1)+\omega(\tA_2)\leq 1,\;\forall\omega\in\Stset$ (\ie transformations that can in
principle occur in the same experiment). The addition of the two coexistent transformations is the
transformation $\tS=\tA_1+\tA_2$ corresponding to the event $e=\{1,2\}$ in which the apparatus
signals that either $\tA_1$ or $\tA_2$ occurred, but does not specify which one. Such transformation
is uniquely determined by the informational and dynamical classes as $\forall\omega\in\Stset$:
$\omega(\tA_1+\tA_2)=\omega(\tA_1)+\omega(\tA_2),\; (\tA_1+\tA_2)\omega=\tA_1\omega+ 
\tA_2\omega$.  The composition "$\circ$" of transformations is distributive with respect to the
addition "$+$". We can also define the multiplication $\lambda\tA$ of a transformation $\tA$ by a
scalar $0\leq\lambda\leq 1$ as the transformation dynamically equivalent to $\tA$, but occurring
with rescaled probability $\omega(\lambda\tA)=\lambda\omega(\tA)$. Now, since for every couple of
transformation $\tA$ and $\tB$ the transformations $\lambda\tA$ and $(1-\lambda)\tB$ are coexistent
for $0\leq\lambda\leq 1$, the set of transformations also becomes a convex set. Moreover, since the
composition $\tA\circ\tB$ of two transformations $\tA$ and $\tB$ is itself a transformation and
there exists the identical transformation $\tI$ satisfying $\tI\circ\tA=\tA\circ\tI=\tA$ for every
transformation $\tA$, the transformations make a semigroup with identity, \ie a {\em monoid}.
Therefore, the set of physical transformations $\Trnset$ is a convex monoid.

It is obvious that we can extend the notions of coexistence, sum and multiplication by a scalar from
transformations to effects via equivalence classes.

\par A purely dynamical notion of {\bf independent systems} coincides with the possibility of
performing local experiments. More precisely, we say that two physical systems are {\em independent}
if on the two systems 1 and 2 we can perform {\em local experiments} $\AA^{(1)}$ and $\AA^{(2)}$
whose transformations commute each other (\ie
$\tA^{(1)}\circ\tB^{(2)}=\tB^{(2)}\circ\tA^{(1)},\;\forall \tA^{(1)}\in\AA^{(1)},\,\forall
\tB^{(2)}\in\AB^{(2)}$).  Notice that the above definition of independent systems is purely
dynamical, in the sense that it does not contain any statistical requirement, such as the existence
of factorized states. Indeed, the present notion of dynamical independence is so minimal that it can
be satisfied not only by the quantum tensor product, but also by the quantum direct sum. As shown in
Ref. \cite{darianoVax2006}, it is an additional Postulate---the {\em local observability
  principle}---which selects the tensor product. In the following, when dealing with more than
one independent system, we will denote local transformations as ordered strings of transformations
as follows $\tA,\tB,\tC,\ldots:=\tA^{(1)}\circ\tB^{(2)}\circ\tC^{(3)}\circ\ldots$. For effects one
has the locality rule $([\tA]_\eff,[\tB_\eff)\in[(\tA,\tB)]_\eff$. The notion of independent systems
now entails the notion of {\em local state}---the equivalent of partial trace in Quantum Mechanics.
For two independent systems in a joint state $\Omega$, we define the {\bf local state} $\Omega|_1$
(and similarly $\Omega|_2$) as the probability rule $\Omega|_1(\tA)\doteq\Omega(\tA,\tI)$ of the
joint state $\Omega$ with a local transformation $\tA$ acting only on system $1$ and with all other
systems untouched.

\section{The C${}^*$-algebra of transformations} 
We have seen that the physical transformations make a convex monoid. It is easy to extend it to a
real algebra $\Trnset_\Reals$ by taking differences of physical transformations, and multiply them
by scalars $\lambda>1$. We will call the elements of $\Trnset_\Reals$ that are not in $\Trnset$ {\em
  generalized transformations}. Likewise, we can introduce generalized effects, and denote their
linear space as $\Cntset_\Reals$. Now that we have a real algebra of generalized transformations and
a linear space of generalized effects we want to introduce a positive bilinear form over them, by
which we will be able to introduce a scalar product via the GNS construction\cite{GNS}. The role of such
bilinear form will be played by a {\em faithful state}. 

\par We say that a state $\Phi$ of a bipartite system is {\bf dynamically faithful} for system 1 
when for every transformation $\tA$ the map $\tA\leftrightarrow(\tA,\tI)\Phi$ is one-to-one. This
means that for every bipartite effect $\cB$ one has
$\Phi(\cB\circ(\tA,\tI))=0\quad\Longleftrightarrow\quad\tA=0$.  Clearly the correspondence remains
one-to-one when extended to $\Trnset_\Reals$. On the other hand, we will call a state $\Phi$ of a
bipartite system {\bf preparationally faithful} for system 1 if every joint bipartite state $\Omega$
can be achieved by a suitable local transformation $\tT_\Omega$ on system 1 occurring with nonzero
probability. Clearly a bipartite state $\Phi$ that is preparationally faithful is also {\em locally}
preparationally faithful, namely every local state $\omega$ of system 2 can be achieved by a
suitable local transformation $\tT_\omega$ on system 1.
\par In Postulate \ref{p:faith} we also use the notion of {\bf symmetric joint state}. This is
simply defined as a joint state of two identical systems such that for any couple of transformations
$\tA$ and $\tB$ one has $\Phi(\tA,\tB)=\Phi(\tB,\tA)$.

Clearly both notions of faithfulness hold for both systems for a symmetrical state. For a {\em
  faithful} bipartite state $\Phi$, the {\em transposed transformation} $\tA'$ of the transformation
$\tA$ is the generalized transformation which when applied to the second component system gives the
same conditioned state and with the same probability as the transformation $\tA$ operating on the
first system, namely $(\tA,\tI)\Phi=(\tI,\tA')\Phi$ or, equivalently
$\Phi(\cB\circ\tA,\cC)=\Phi(\cB,\cC\circ\tA')$ $\forall \cB,\cC\in\Cntset$.  Clearly the transposed
is unique, due to injectivity of the map $\tA\leftrightarrow(\tA,\tI)\Phi$, and it is easy to check
the axioms of transposition ($(\tA+\tB)'= \tA'+\tB'$, $(\tA')'=\tA$, $(\tA\circ\tB)'=
\tB'\circ\tA'$) and that $\tI'=\tI$.

The main ingredient of a GNS construction for representing transformations would be a positive form
$\varphi$ by which one can construct a scalar product as $\<\tA|\tB\>:=\varphi(\tA^\dag\circ\tB)$,
in terms of which we then have
$\<\tA|\tC\circ\tB\>=\<\tC^\dag\circ\tA|\tB\>\equiv\varphi(\tA^\dag\circ\tC\circ\tB)
=\varphi((\tC^\dag\circ\tA)^\dag\circ\tB)$. However, we don't have a definition for the adjoint, and
it is not easy to devise a positive form over generalized transformations $\Trnset_\Reals$ such that
the transposition plays the role of the adjoint on a real Hilbert space. Indeed, if we take
$\varphi$ as the local state of a symmetric faithful state $\varphi=\Phi|_2\equiv\Phi|_1$ we have
$\varphi(\tA'\circ\tB)=\Phi(\tA',\tB')\equiv\Phi(\cA',\cB')$ (notice that the bilinear form $\Phi$
is actually defined on effects), but the fact that $\Phi$ is positive over the convex set $\Trnset$
of physical transformations doesn't guarantee that its extension to generalized transformations
$\Trnset_\Reals$ is still positive. One can, however, extract from $\Phi$ a positive bilinear form
over $\Cntset_\Reals$ in terms of its absolute value $|\Phi|:=\Phi_+-\Phi_-$. Indeed, the absolute
value can be defined thanks to the fact that $\Phi$ is real symmetric, whence it can be diagonalized
over $\Cntset_\Reals$. Upon denoting by $\map{P}_\pm$ the orthogonal projectors over the linear
space corresponding to positive and negative eigenvalues, respectively, \footnote{The existence of
  the orthogonal space decomposition corresponding to positive and negative eigenvalues is
  guaranteed for finite dimensions. For infinite dimensions $\Phi$  is just a symmetric form over a
  Banach space, and the existence of such decomposition remains to be seen.} one has
$|\Phi|(\cA,\cB)=\Phi(\varsigma(\cA),\cB)$, where $\varsigma(\cA):=(\map{P}_+-\map{P}_-)(\cA)$.  The
map $\varsigma$ is an involution, namely $\varsigma^2=\map{I}$. The fact that the state is also
preparationally faithful implies that the bilinear form is {\em strictly}
positive\cite{darianoVax2006} (namely $|\Phi|(\cC,\cC)=0$ implies that $\cC=0$). We can extend the
involution $\varsigma$ to generalized transformations by considering the absolute value of $\Phi$
regarded as bilinear form over generalized transformations $\Trnset_\Reals$. In this way we have
$\varphi(\varsigma(\tA')\circ\tB)$ as a positive form, and we can identify
$\varsigma(\tA')\equiv\tA^\dag$ as the adjoint, namely as the composition of the transposition and
the {\em complex conjugation} $\varsigma$. We need to choose the extension of $\varsigma$ to
transposition to be composition-preserving, \ie
$\varsigma(\tB\circ\tA)=\tB^\varsigma\circ\tA^\varsigma$,\footnote{The involution $\varsigma$ is
  composition-preserving if $\varsigma(\Trnset)=\Trnset$ namely if the involution preserves physical
  transformations.  Indeed, for such an involution one can consider its action on transformations
  induced by the involutive isomorphism $\omega\to\omega^\varsigma$ of the convex set of states
  $\Stset$ defined as $\omega^\varsigma(\tA):=\omega(\varsigma(\tA))$,
  $\forall\omega\in\Stset,\;\forall\tA\in\Trnset$. Consistency with state-reduction
  $\omega_\tA^\varsigma(\tB)\equiv\omega_{\tA^\varsigma}(\tB^\varsigma)$
  $\forall\omega\in\Stset,\;\forall\tA,\tB\in\Trnset$ is then equivalent to
  $\omega(\varsigma(\tB\circ\tA))=\omega(\tB^\varsigma\circ\tA^\varsigma)$
  $\forall\omega\in\Stset,\;\forall\tA,\tB\in\Trnset$. The involution $\varsigma$ of $\Stset$ is
  just the inversion of the principal axes corresponding to negative eigenvalues of the symmetric
  bilinear form $\Phi$.\cite{darianoVax2006})} in such a way that
$(\tB\circ\tA)^\dag=\tA^\dag\circ\tB^\dag$ (\ie it is the transposition that takes care of ordering).
Now, following the GNS construct, we introduce the scalar product as
${}_\Phi\!\<\cA|\cB\>_\Phi:=\varphi(\tA^\dag\circ\tB)= \Phi(\varsigma(\cA'),\cB')$, and we can
verify that $\tA^\dag:=\varsigma(\tA')$ works as an adjoint for such scalar product, namely
${}_\Phi\!\<\tC^\dag\circ\cA|\cB\>_\Phi={}_\Phi\!\<\cA|\tC\circ\cB\>_\Phi$.\footnote{Clearly in this
  way one recovers the customary operator-like action of transformations from the left
  $|\underline{\tC\circ\tA}\>_\Phi= |\tC\circ\cA\>_\Phi$ which follows from the fact that the scalar
  product is defined in terms of the positive bilinear form $|\Phi|$ over transposed transformations
  ${}_\Phi\!\<\tC\circ\cA|\tB\>_\Phi=|\Phi|(\cA'\circ\tC',\cB')=\Phi(\varsigma(\cA'\circ\tC'),\cB')$.}
In the following we will equivalently write the entries of the scalar product as generalized
transformations or as generalized effects, with
${}_\Phi\!\<\tA|\tB\>_\Phi:={}_\Phi\!\<\cA|\cB\>_\Phi$, the generalized effects being the actual
vectors of the linear factor space of generalized transformations modulo informational equivalence.

Now, by taking complex linear combinations of generalized transformations and defining
$\varsigma(c\tA)=c^*\varsigma(\tA)$ for $c\in\Cmplx$, we can extend the adjoint to complex linear
combinations of generalized transformations, whose linear space will be denoted by $\Trnset_\Cmplx$,
which is a complex algebra that we will also denote as $\aA$. On the other hand, we can trivially
extend the real pre-Hilbert space of generalized effects $\Cntset_\Reals$ to a complex pre-Hilbert
space $\Cntset_\Cmplx$ by just considering complex linear combinations of generalized effects.

The remaining setting up of the C${}^*$-algebra representation of $\aA$ is just standard GNS
construction.  We now have a scalar product ${}_\Phi\!\<\tA|\tB\>_\Phi=\Phi_2(\tA^\dag\circ\tB)$
between transformations. Symmetry and positivity imply the bounding\cite{darianoVax2006}
${}_\Phi\!\<\tA|\tB\>_\Phi\leq\n{\tA}_\Phi\n{\tB}_\Phi$, where we introduced the norm induced by the
scalar product $\n{\tA}_\Phi^2\doteq{}_\Phi\!\<\tA|\tA\>_\Phi$.  By taking the equivalence classes
$\aA/\aI$ with respect to the zero-norm elements $\aI\subseteq\aA$ we thus obtain a complex
pre-Hilbert space equipped with a symmetric scalar product, and, since the scalar product is
strictly positive over generalized effects, the elements of $\aA/\aI$ are indeed the generalized
effects, \ie $\aA/\aI\simeq\Cntset_\Cmplx$ as linear spaces.  Moreover, from the bounding for the
scalar product it follows that the set $\aI\subseteq\aA$ of zero norm elements $\tX\in\aA$ is a left
ideal (\ie $\tX\in\aI$, $\tA\in\aA$ implies $\tA\circ\tX\in\aI$), whence using our scalar product
defined as ${}_\Phi\!\<\tA|\tB\>_\Phi=\Phi_2(\tA^\dag\circ\tB)$ we can represent elements of $\aA$
($\aA\equiv\Trnset_\Cmplx$ are the generalized complex transformations) as operators over the
pre-Hilbert space of effects and make $\aA$ a C${}^*$-algebra.  We just need to introduce the norm
on transformations as $\n{\tA}_\Phi:=\sup_{\cB\in\Cntset_\Cmplx,\n{\cB}_\Phi\leq
  1}\n{\tA\circ\cB}_\Phi$. Completion of $\aA/\aI\simeq\Cntset_\Cmplx$ in the norm topology will
make it a Hilbert space that we will denote by $\sH_\Phi$.  Such completion also implies that
$\Trnset_\Cmplx\simeq\aA$ can be completed to a complex C$^*$-algebra (\ie a Banach algebra
satisfying the identity $\n{\tA^\dag\circ\tA}=\n{\tA}^2$), as it can be easily proved by standard
techniques\cite{darianoVax2006}. 

The product in $\aA$ defines the action of $\aA$ on the vectors in $\aA/\aI$, by associating to each
element $\tA\in\aA$ the linear operator $\pi_\Phi(\tA)$ defined on the dense domain
$\aA/\aI\subseteq\sH_\Phi$ as $\pi_\Phi(\tA)|\cB\>_\Phi\doteq|\underline{\tA\circ\tB}\>_\Phi$.  The
fact that $\aA$ is a Banach algebra also implies that the domain of definition of $\pi_\Phi(\tA)$
can be easily extended to the whole $\sH_\Phi$ by continuity.  From the definition of the scalar
product, and using the fact that the state $\Phi$ is also preparationally faithful according to
Postulate \ref{p:faith}, the Born rule can be written in the GNS representation as
$\omega(\cA)={}_\Phi\<\underline{\tA^\dag}|\varrho\>_\Phi$, with representation of state
$\varrho=\cT_\omega'/\Phi(\cT_\omega,\tI)$ \cite{darianoVax2006}, $\cT_\omega$ denoting the
transformation on system 2 corresponding to the local state $\omega$ on system 1.  Then, the
representation of transformations is
$\omega(\cB\circ\tA)={}_\Phi\<\underline{\tB^\dag}|\tA|\rho\>_\Phi:=
{}_\Phi\<\underline{\tB^\dag}|\tA\circ\rho\>_\Phi\equiv
{}_\Phi\<\tA^\dag\circ\underline{\tB^\dag}|\rho\>_\Phi$.

\ack I acknowledge illuminating discussions with M. Ozawa. This work has been supported by Ministero
Italiano dell'Universit\`a e della Ricerca (MIUR) through PRIN 2005.


\begin{thebibliography}{9}
\itemsep 0pt
\bibitem{darianoVax2006} G.~M. D'Ariano, {\em Operational Axioms for Quantum Mechanics} in
  \emph{Quantum Theory, Reconsideration of Foundations - 4}, ed. by G.~Denier, A.~Y. Khrennikov, and
  T.~M.  Nieuwenhuizen (AIP, Melville, New York, 2007 in press) [also \textbf{quant-ph/0611094}].
\bibitem{dariano-beyond} G. M. D'Ariano, {\em Where the mathematical structure of Quantum Mechanics
    comes from}, in {\em Beyond the Quantum} (World Scientific in press) [also
  \textbf{quant-ph/0612162}] (see also references to previous papers here).
\bibitem{tomo_lecture} G.~M. D'Ariano, \emph{Tomographic methods for universal estimation in quantum
    optics}, IOS Press, Amsterdam, 2002, pp. 385--406, scuola ``E. Fermi'' on {\em Experimental
    Quantum Computation and Information}.
\bibitem{calib} G.~M. D'Ariano, P.~L. Presti, and L.~Maccone, \emph{Phys. Rev. Lett.}  \textbf{93},
  250407 (2004); G.~M. D'Ariano, and P.~L. Presti, \emph{Phys. Rev. Lett.} \textbf{86}, 4195 (2001);
  G.~M. D'Ariano, and P.~L. Presti, \emph{Phys. Rev. Lett.} \textbf{91}, 047902--1--4 (2003).
\bibitem{GNS} I.~M. Gelfand, and M.~A. Neumark, \emph{Mat. Sb.} \textbf{12}, 197 (1943).
\bibitem{Ludwig-axI} G.~Ludwig, \emph{An Axiomatic Basis for Quantum Mechanics I: Derivation of
    Hilbert Space Structure}, Springer, SPR:adr, 1985.
\end{thebibliography}
\end{document}